\documentclass[conference,a4paper]{IEEEtran}
\usepackage{xcolor}
\usepackage{balance}

\ifCLASSINFOpdf
  \usepackage[pdftex]{graphicx}
\else
  \usepackage[dvips]{graphicx}
\fi

\ifCLASSOPTIONcompsoc
 \usepackage[caption=false,font=normalsize,labelfont=sf,textfont=sf]{subfig}
\else
 \usepackage[caption=false,font=footnotesize]{subfig}
\fi

\usepackage{amsmath}

\begin{document}

\title{Time Interfaces in Nanoplasma-Switched Wire Media}

\author{\IEEEauthorblockN{
Mikhail Sidorenko\IEEEauthorrefmark{1},    
Sergei Tretyakov\IEEEauthorrefmark{1},
 and  Constantin Simovski\IEEEauthorrefmark{1}
}                                     

\IEEEauthorblockA{\IEEEauthorrefmark{1}
Department of Electronics and Nanoengineering, Aalto University, Espoo, Finland}  
 \IEEEauthorblockA{ \emph{mikhail.sidorenko@aalto.fi} }
}

\maketitle

\begin{abstract}
In this work, we consider instantaneous transitions of an infinitely extended uniaxial dielectric into a wire medium (WM) of continuous infinitely long conducting wires. Due to the strong spatial dispersion in the WM the known (Morgenthaler's) theory of temporal discontinuities is not applicable. We solve this problem analytically in time domain. We show that a transverse electromagnetic (TM) plane  wave  transforms into four waves: a pair of TM waves and a pair of transverse electromagnetic  waves. This way, the power flow splits into two different directions, with one of them along the wires. Such a transition can possibly be achieved by nanoplasma discharges in the gaps of the split wires, initiated by an external voltage source applied to the wire and transforming the split wires forming the uniaxial dielectric into continuous ones. 
\end{abstract}

\vskip0.5\baselineskip

\section{Introduction}

Time-varying materials have recently attracted strong attention due to their potential for advanced wave manipulations. Time modulations of metamaterials  enabled new ways for realisations of magnet-less non-reciprocity, wave amplification, frequency conversion, pulse shaping, etc. An important case of time modulations is the case of temporal discontinuities, when the electromagnetic properties of the medium undergo abrupt changes on certain moments of time -- practically, during the time much shorter than one cycle of the wave propagating in the medium. Such time interfaces in lossless and dispersion-free media were considered in 1958 by F. Morgenthaler \cite{Morgentaller}. His theory shows that a monochromatic plane wave transforms into two new waves: the time-refracted one, propagating in the same direction  as the initial wave, and the time-reflected one, propagating in the  opposite direction. In fact, this theory dictates the conservation of the 
wavevector, and the time-reflected wave (effectively having a positive frequency) mathematically corresponds to the solution with the same wavevector and a negative frequency. The Morgenthaler theory does not cover the case when the medium before the discontinuity or after it has spatial dispersion. In particular, this is the case of the so-called wire medium (WM) whose spatial dispersion is resonant at all frequencies \cite{Belov}. In this article, we present an analytical model of temporal discontinuities that transform a uniaxial dielectric (such as formed by thin conducting wires of small (non-resonant) length) into a WM. An exact analytical solution is obtained beyond the Morgenthaler theory, solving an initial value problem in time domain. We propose to implement such transitions in the THz range of frequencies via the so-called nanoplasma discharge \cite{Nano1,Nano2} arising in micron or submicron gaps between sections of metal wires. This discharge may be induced by external DC voltages applied to every wire in a collinear array. 

\section{Solid Wire Medium and Split-Wire Medium}

\begin{figure}[b]
\begin{center}
\includegraphics[width=0.4\textwidth]{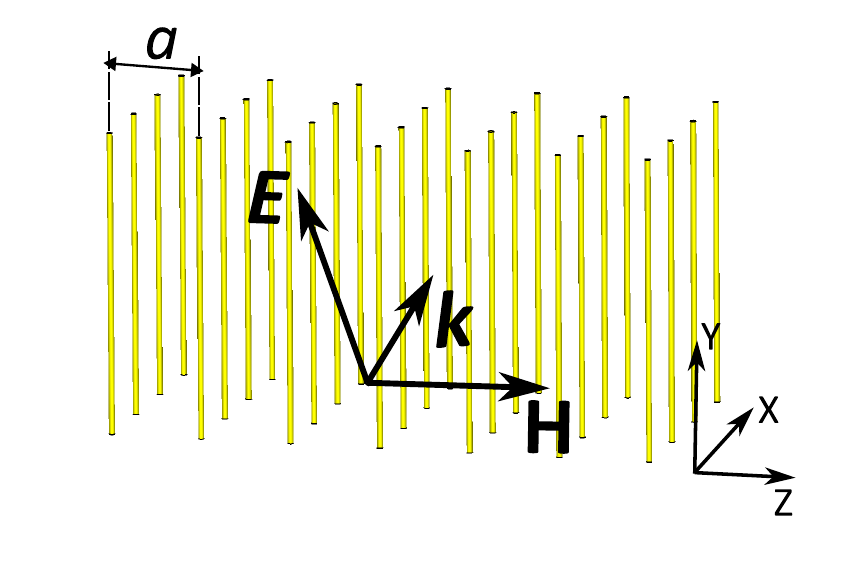}
\caption{\label{Fig1} A schematic view of a wire medium.}
\end{center}
\end{figure}

\begin{figure}[b]
\begin{center}
\includegraphics[width=0.4\textwidth]{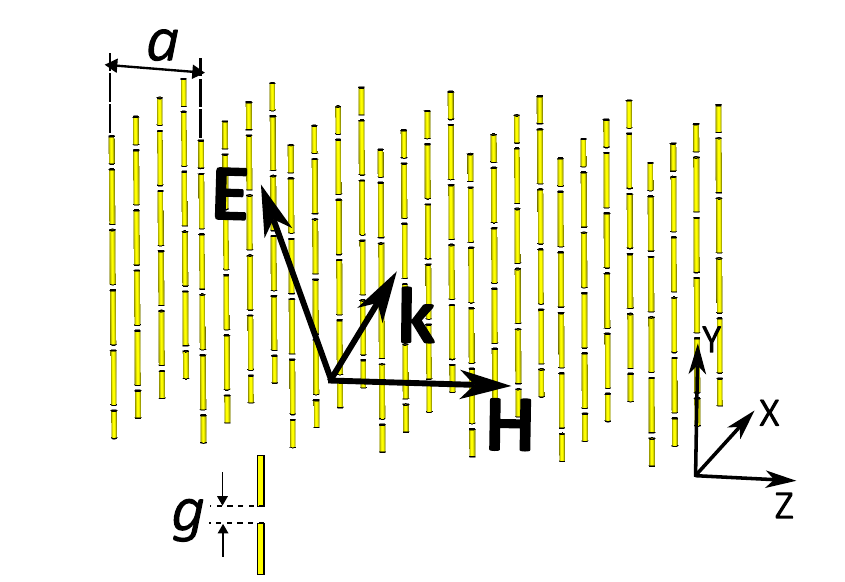}
\caption{\label{Fig2} A schematic view of a split wire medium.}
\end{center}
\end{figure}

Usually, samples of WM are realized as square arrays of thin parallel wires, as is schematically depicted in Fig.~\ref{Fig1}. The wire length is electrically large and can be assumed to be infinite. The period $a$ is much smaller than the free-space wavelength $\lambda_0$, and the wire radius $r_0$ is much smaller than $a$. An overview of electromagnetic properties of such media can be found, for instance, in \cite{Simovski}. The electromagnetic response is described by the tensor of permittivity which essentially depends on the axial ($y-$)component of the wavevector $\mathbf{k}$, i.e., it is strongly spatially dispersive: 
\begin{equation}\label{EpsWire}
    \widetilde{\varepsilon}=\left(\begin{array}{ccc}
        1 & 0 & 0 \\
        0 & \varepsilon_{yy} & 0 \\
        0 & 0 & 1
    \end{array} \right), \qquad
    \varepsilon_{yy}=1-\frac{k_p^2}{k_0^2-k_y^2}.
\end{equation}
Here,  $k_0=\omega_0/c=\omega_0\sqrt{\varepsilon_0\mu_0}$ is the free-space wavenumber, and $k_p=\omega_p/c=\omega_p\sqrt{\varepsilon_0\mu_0}$ is called the WM plasma wavenumber, where $\omega_p$ is the WM plasma frequency, depending in a known  way on the geometrical parameters $a$ and $r_0$, see e.g. in \cite{Simovski}.

As it is explained in \cite{Simovski}, similar arrays of finite wires can be considered as  slabs of WM whose properties inside the slab correspond to the WM of infinitely long wires, i.e. are described by formula \ref{EpsWire}. Special boundary conditions are introduced at the ends of the wires (slab interfaces). 
However, a lattice that is schematically depicted in Fig. \ref{Fig2} -- the lattice of finite metal rods whose length $l$, period $a$ in the $xz$-plane, and period $d=l+g$ along the axis $y$ ($g$ is the gap between the adjacent rod ends) are all small compared to $\lambda_0$ -- should not be considered as a multilayer structure of WM slabs. The electromagnetic model of such lattices is much simpler -- it is a uniaxial dielectric whose spatial dispersion is negligibly small (see e.g. in \cite{Collin}): 
\begin{equation}
    \widetilde{\varepsilon}=\left(\begin{array}{ccc}
        1 & 0 & 0 \\
        0 & \varepsilon_1 & 0 \\
        0 & 0 & 1
    \end{array} \right),\label{f2}
\end{equation}
where $\varepsilon_1$ is a number depending in a known way (see e.g. in \cite{Collin}) on the geometrical parameters $l,a,d,r_0$. 
Below, this lattice is called the split-wire medium (SWM). 

Let a plane wave of frequency $\omega_1<\omega_p$ whose magnetic field $\mathbf{H}=(0,0,H_z)$ is polarized along $z$ propagate in the space filled with the SWM. The wavevector $\mathbf{k}=(k_x,k_y,0)$ and the electric field $\mathbf{E}=(E_x, E_y, 0)$  lie  in the plane $xy$. Using Maxwell's equations and formula \eqref{f2}, the field components can be easily expressed through the  amplitude $H_0$ and the axial permittivity $\varepsilon_1$, and the frequency can be related with the wave vector:
\begin{equation}
    H_z=H_0 \cos\left(\mathbf{k}\cdot \mathbf{r} - \omega_1 t\right), \qquad \omega_1=c\sqrt{\frac{k_x^2}{\varepsilon_1}+k_y^2},
\end{equation}
\begin{equation}\label{SolUniax}
    \begin{array}{l}
         \displaystyle E_x = -H_0 \frac{k_y c}{\omega_1} \sqrt{\frac{\mu_0}{\varepsilon_0}} \cos\left(\mathbf{k}\cdot \mathbf{r} - \omega_1 t \right),  \\ \\
         \displaystyle E_y = H_0 \frac{k_x c}{\omega_1} \frac{1}{\varepsilon_1} \sqrt{\frac{\mu_0}{\varepsilon_0}} \cos\left(\mathbf{k}\cdot \mathbf{r} - \omega_1 t\right). 
    \end{array}
\end{equation}
Let us learn what happens with this wave if the SWM instantaneously transits into a WM.

\section{Transition between Split-Wire Medium and Wire-Medium}
\subsection{Governing Equations}

Let the transition occurs at the moment $t=0$ (the transition time is much smaller than  $2\pi/\omega_1$). In the wires of the WM the electric currents $I(x,y,z)$ arise which depend in a discrete way on the $x$ and $z$ coordinates of the wires and in the model of the effectively continuous medium can be described via the volumetric current density  $\mathbf{J}=(0, J_y, 0)$ ($J_y=I/a^2$). Since the continuity equation $\mathbf{J}=-\partial \mathbf{P}/\partial t$ relates the current density with the bulk polarization density vector $\mathbf{P}$, having  only 
$y$-component (the model of thin wires), the electromagnetic response of the WM can be described by $P_y\equiv P$. Maxwell's equations for the electromagnetic field and polarization read as 
\begin{equation}\label{Maxwell2}
    \left\{ \begin{array}{l} \displaystyle
         \frac{\partial E_y}{\partial x} - \frac{\partial E_x}{\partial y}  = -\mu_0 \frac{\partial H_z}{\partial t} \\ \\
         \displaystyle \frac{\partial H_z}{\partial x} = -\varepsilon_0 \frac{\partial E_y}{\partial t} - \frac{\partial P}{\partial t} \\ \\
         \displaystyle \frac{\partial H_z}{\partial y}  = \varepsilon_0 \frac{\partial E_x}{\partial t} 
    \end{array} \right. .
\end{equation}
After some algebra these equations give:
\begin{equation}
    \frac{\partial^2 E_y}{\partial x^2} + \frac{\partial^2 E_y}{\partial y^2} - \frac{1}{c^2} \frac{\partial^2 E_y}{\partial t^2} = \mu_0 \frac{\partial^2 P}{\partial t^2} - \frac{1}{\varepsilon_0} \frac{\partial^2 P}{\partial y^2}.
\end{equation}
This first governing equation is valid for both WM and SWM, it only uses the fact that the polarization of the medium is purely electric and is directed along $y$.   

To obtain the second governing equation we may recall the definitions of the displacement vector $\mathbf{D}$ and the effective permittivity:
\begin{equation}
    D_y = \varepsilon_0 E_y + P = \varepsilon_0 \varepsilon_{yy} E_y
\end{equation}
and consider the relation $P=(\varepsilon_{yy}-1)E_y$ in the limit case $t\rightarrow\infty$. Then we may present this relation in the Fourier domain and involve formula (\ref{EpsWire}) that results in the following equation: 
\begin{equation}
        (k^2-k_y^2) \widehat{P} = -\varepsilon_0 k_p^2 \widehat{E}_y.
\end{equation}
In the space-time domain the counterpart of the factor $k^2$ is the differential operator $-\partial^2/c^2\partial t^2$, and the counterpart of $k_y^2$ is $-\partial^2/\partial y^2 $. Then the equations on $E_y$ and $P$ form the following system:
\begin{equation}\label{MainSystemOfEq}
    \left\{ \begin{array}{l}
        \displaystyle \frac{\partial^2 P}{\partial y^2} - \frac{1}{c^2} \frac{\partial^2 P}{\partial t^2} = -\varepsilon_0 k_p^2 E_y,\\ \\
        \displaystyle \Delta E_y - \frac{1}{c^2} \frac{\partial^2 E_y}{\partial t^2} = k_p^2 E_y .
    \end{array} \right.
\end{equation}
This system of equations describes the polarization properties of the WM in the time domain. Since these properties do not change after the temporal discontinuity $t=0$, equations (\ref{MainSystemOfEq}) are valid not only for $t\rightarrow\infty$, they are valid starting from the moment $t=0$.  

\subsection{Initial-Value Problem and Solution}

To find the electromagnetic field after the transition of the media at $t=0$ we pose the conditions that formulate the initial-value problem for the system of governing equations (\ref{MainSystemOfEq}):
\begin{equation}\label{InitialValueProblem1}
    \left\{ \begin{array}{l}
        \displaystyle P|_{t=0} = \varepsilon_0(\varepsilon_1-1) E_0 \cos(\mathbf{k}\cdot \mathbf{r}), \\   \\
        \left. \displaystyle \frac{\partial P}{\partial t}\right|_{t=0} = \varepsilon_0 (\varepsilon_1 - 1) \omega_1 E_0 \sin(\mathbf{k} \cdot \mathbf{r}), \\  \\
         \displaystyle E_y|_{t=0} = E_0\cos (\mathbf{k} \cdot \mathbf{r}), 
         \\  \\ \left. \displaystyle \frac{\partial E_y}{\partial t}\right|_{t=0} = \omega_1 E_0\sin (\mathbf{k} \cdot \mathbf{r}) 
    \end{array} \right. 
\end{equation}
These conditions express the continuity of $E_y$, of $\partial E_y/\partial t$, of $P$ and of $J=-\partial P/\partial t$ at $t=0$. The continuity of $J$ follows from the non-zero inductance of the wire medium (the jump of $J$ would mean the infinite energy). The continuity of $\partial E_y/\partial t$ is not so evident. We have proved this continuity from  causality of the polarization response in the time domain and the symmetry of the time-domain Green function $G(\mathbf{r},t)$. As to the time-continuity of $E_y$ and $P$, it follows from the existence and continuity of their time derivatives.

The electromagnetic field that satisfies the governing equations (\ref{MainSystemOfEq}) and the initial-value problem was found using the Laplace transform, and it has the following components:
\begin{equation}\label{AllResultsEx}
    \begin{array}{r}
        \displaystyle E_x(\mathbf{r},t) = \frac{-k_y}{k_x \varepsilon_0} \left\{ \left[\frac{\varepsilon_0 E_0}{2} \left(1 + \frac{\omega_1}{\omega_2} \right) + A^+ \right] \cos(\mathbf{k}\cdot \mathbf{r} - \omega_2 t) \right. \\ \\
        \displaystyle \left. +  \left[\frac{\varepsilon_0 E_0}{2} \left(1 - \frac{\omega_1}{\omega_2} \right) + A^- \right] \cos(\mathbf{k}\cdot \mathbf{r} + \omega_2 t) \right\} \\ \\
        \displaystyle +  B^+ \cos(\mathbf{k}\cdot \mathbf{r} - \omega_2^* t) + B^- \cos(\mathbf{k}\cdot \mathbf{r} + \omega_2^* t),
    \end{array}
\end{equation}
\begin{equation}\label{AllResultsEy}
    \begin{array}{r}
         \displaystyle E_y(\mathbf{r},t) = \frac{E_0}{2} \left(1 + \frac{\omega_1}{\omega_2} \right) \cos(\mathbf{k} \cdot \mathbf{r} - \omega_2 t) \\ \\ 
         \displaystyle + \frac{E_0}{2} \left(1 - \frac{\omega_1}{\omega_2} \right) \cos(\mathbf{k} \cdot \mathbf{r} + \omega_2 t),
    \end{array}
\end{equation}
\begin{equation}\label{AllResultsHz}
    \begin{array}{r}
         \displaystyle H_z(\mathbf{r},t) = \frac{\omega_2}{k_x} \left\{ \left[\frac{\varepsilon_0 E_0}{2} \left(1 + \frac{\omega_1}{\omega_2} \right) + A^+ \right] \cos(\mathbf{k}\cdot \mathbf{r} - \omega_2 t) \right. \\ \\
         \displaystyle \left. -  \left[\frac{\varepsilon_0 E_0}{2} \left(1 - \frac{\omega_1}{\omega_2} \right) + A^- \right] \cos(\mathbf{k}\cdot \mathbf{r} + \omega_2 t) \right\} \\ \\
         \displaystyle + \frac{\omega_2^*}{k_x} \left[  B^+ \cos(\mathbf{k}\cdot \mathbf{r} - \omega_2^* t) -  B^- \cos(\mathbf{k}\cdot \mathbf{r} + \omega_2^* t)  \right],
    \end{array}
\end{equation}
\begin{equation}    \label{AllResultsP}
\begin{array}{r}
    P(\mathbf{r},t) = A^+\cos(\mathbf{k}\cdot \mathbf{r} - \omega_2 t) + A^- \cos(\mathbf{k}\cdot \mathbf{r} + \omega_2 t)   \\ \\
    + B^+\cos(\mathbf{k}\cdot \mathbf{r} - \omega_2^* t) + B^- \cos(\mathbf{k}\cdot \mathbf{r} + \omega_2^* t),
\end{array}
\end{equation}
where
\begin{equation}\label{AllResultsCoeffs}
    \begin{array}{r}
        \displaystyle A^{\pm}=\frac{E_0}{2} \frac{\varepsilon_0 k_p^2}{k_x^2+k_p^2} \left(1 \pm \frac{\omega_1}{\omega_2} \right),   \\   \displaystyle \omega_2 = c\sqrt{k_x^2 + k_y^2 + k_p^2}, \qquad \omega_2^*=k_y c, \\ \\ 
        \displaystyle B^{\pm}=-\frac{E_0}{4} \frac{\varepsilon_0 k_p^2}{k_x^2+k_p^2} \left[\left(1+\frac{\omega_1}{\omega_2}\right) \left(1\pm\frac{\omega_2}{\omega_2^*}\right) \right. \\ \\
        \displaystyle + \left. \left(1-\frac{\omega_1}{\omega_2}\right) \left(1\mp\frac{\omega_2}{\omega_2^*}\right) \right] + \frac{\varepsilon_0 (\varepsilon_1-1)E_0}{2} \left(1 \pm \frac{\omega_1}{\omega_2^*} \right).  
    \end{array}
\end{equation}
Since the initial frequency $\omega_1$ is related with the wavevector components as 
$\omega_1 = c\sqrt{{k_x^2}/{\varepsilon_1} + k_y^2}$ (see above), one can easily express the new frequencies $\omega_2$ and $\omega_2^*$ through $\omega_1$. These new frequencies  
are discussed below. 

\subsection{Analysis of the Formulas}

From these formulas it is clear that after the time moment $t=0$, a frequency conversion takes place -- instead of the initial frequency $\omega_1$ two new frequencies $\omega_2$ and $\omega_2^*$ arise. These two frequencies correspond to two different types of waves. Recall that the initial frequency $\omega_1$ corresponds to a TM wave whose energy propagates under certain angles  to the  axes $x$ and $y$. The same propagation angle corresponds also to two waves with the new frequency $\omega_2$. 
This frequency can be expressed via $\omega_1$ and the angle $\psi$ between the wavevector $\mathbf{k}$ and the axis $x$:
\begin{equation}
    \omega_2 = \sqrt{\omega_1^2\varepsilon_1{1+\tan^2\psi\over 1+\varepsilon_1\tan^2\psi}+c^2k^2_p}.
\end{equation}
Since $\varepsilon_1>0$, this frequency is above the plasma frequency $\omega_p$ of the newly born WM. Waves with such frequencies can propagate in the WM in any direction. Respectively, both waves with the frequency $\omega_2$ carry energy along the same line as the initial wave and keep the same TM polarization. One wave is time-refracted, and its propagation direction coincides with that of the initial wave. Another wave is time-reflected, and its propagation direction is opposite to that of the initial wave. For both these waves, the group velocity can be expressed via the wavevector components in the  following way:
\begin{equation}
    \mathbf{v}_g^{(TM)}=\pm \nabla_k \omega_2 = \pm \frac{c}{\sqrt{k_x^2+k_y^2+k_p^2}} \left( k_x, k_y, 0\right), 
\end{equation}
where plus corresponds to the time-refracted wave and minus to the time-reflected wave.  
This pair of TM waves has the electric field vectors with $\left(E_x^{(1\pm)}, E_y, 0\right)$ components, and the magnetic field vector with the components $\left(0, 0, H_z^{(1\pm)} \right)$. The expressions for these components of the electromagnetic field in the WM are given above. 

The second new frequency, $\omega_2^*$ is below the plasma frequency of the newly born WM because $\omega_1$ is assumed to be below it:
\begin{equation}
    \omega_2^* = \omega_1\sqrt{\varepsilon_1{\tan^2\psi\over 1+\varepsilon_1\tan^2\psi}}<\omega_1.
\end{equation}
If $\omega_1>\omega_p$, the lattice of rods is not a uniaxial dielectric with dispersion-free permittivity $\varepsilon_1$. It is a photonic crystal, and $\omega_1$ belongs to one of high-frequency pass-bands of this photonic crystal. Indeed, our model is not valid in this case, which needs to be studied separately. In our case $\omega_1<\omega_p$, and the new frequency $\omega_2^*$ corresponds to a pair of nonuniform TEM waves. For this wave type, the time-reflected wave is also present, it vanishes only if the initial wave propagated along $y$. The time-refracted and time-reflected TEM waves have the electric field with the components $\left( E_x^{(2\pm)},0,0 \right)$ and the magnetic field $\left(0, 0, H_z^{(2\pm)} \right)$. For these waves, the expression for the group velocity reads as
\begin{equation}
    \mathbf{v}_g^{(\rm TEM)} = \pm \nabla_k \omega_2^* = (0, \pm c, 0).
\end{equation}
So, the energy of both TEM waves propagates along the $y$ axis -- the velocity of this energy transfer does not depend on the direction of the initial wavevector $\mathbf{k}$. These waves result from strong far-field interactions in the array of continuous wires, i.e. from the spatial dispersion of the WM.

The total electromagnetic field at any point of the WM has the components $E_y^{(1+)}+E_y^{(1-)}=E_y$, $E_x^{(1+)}+E_x^{(1-)}+E_x^{(2+)}+E_x^{(2-)}=E_x$, $H_z^{(1+)}+H_z^{(1-)}+H_z^{(2+)}+H_z^{(2-)}= H_z$, as follows from the formulas above.  

\section{Nanoplasma Switching Between Two Media}

A practical implementation of the media transition process  considered above can be based on a gas discharge. We propose to implement such a media transition by using a nanoplasma discharge \cite{Nano1,Nano2}. A nanoplasma discharge is a discharge in a very small gap of the size less than a few micrometers. In this case, the effect of field emission electron dominates over the Townsend discharge mechanism. This leads to a very fast nanoplasma evolution.

\begin{figure}[!h]
\begin{center}
\includegraphics[width=0.4\textwidth]{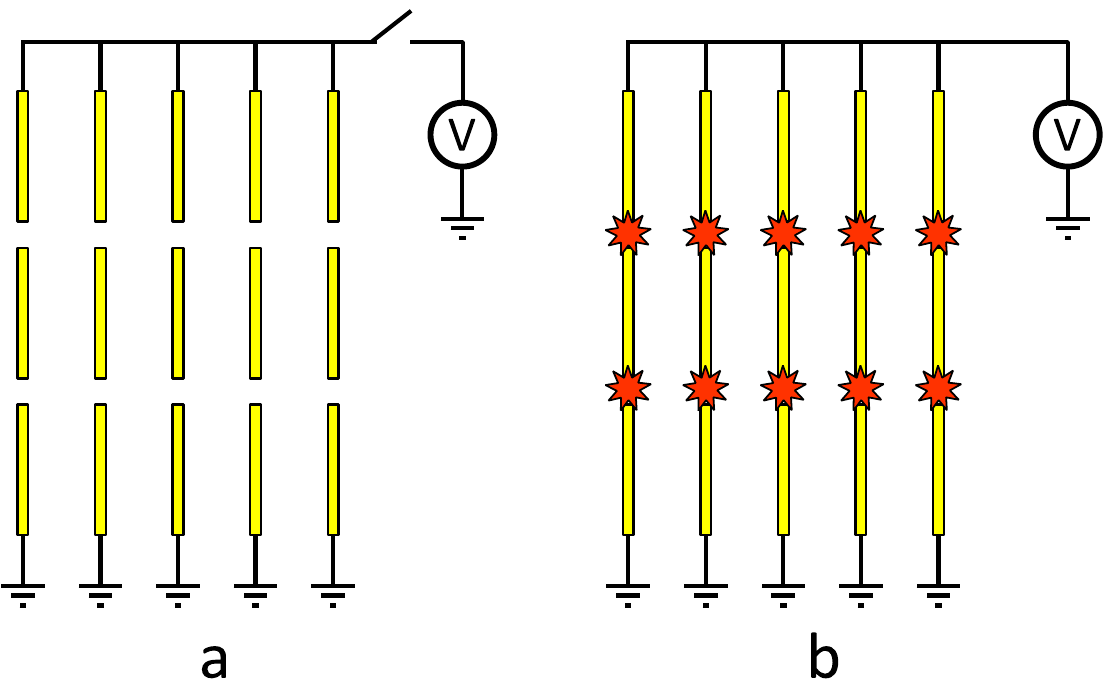}
\caption{\label{Fig3} A schematic of a possible switching between split-wire and wire media: a) when the external voltage source is not applied, the split-wire media exhibits the properties of a uniaxial dielectric, and b) when the external voltage is applied and nanoplasma discharges are ignited, the split-wire media effectively behaves as a spatially dispersive wire media. }
\end{center}
\end{figure}

A schematic of possible switching between split-wire and wire media is shown in Fig.~\ref{Fig3}. An external power source connected to the wires provides high voltage to ignite nanoplasma discharges. The low impedance of the nanoplasma discharges allows the split wires with discharges in the gaps to be efficiently considered as solid conducting wires. Due to the exceptionally fast nanoplasma switch-on process of the order of a few picoseconds, as was reported in \cite{Nano1}, it is theoretically possible to obtain the time interface which was considered in this work up to the sub-terahertz band. The impedance of the gap filled with nanoplasma referred to the period $d$ of the gaps is negligibly small compared to the inductive impedance of the WM per unit lengths. In any case, this impedance per unit length of the WM is inductive. In our estimations, the nanoplasma discharges practically do not change the polarization properties of the WM loaded by this impedance compared to the polarization properties of the conventional WM.

\section{Conclusion}
In this article we have studied a time interface between a uniaxial dielectric and a medium with strong spatial dispersion. The analytical results have been obtained by introducing original time boundary conditions for the electric field, polarisation, and their time derivatives. We have obtained exact analytical results predicting the existence of four waves after the time interface (including two time-refracted waves whose energy propagates in two different direction). The considered media transition can be realised in an artificial media which we call here a 'split-wire media' by ultrafast nanoplasma discharges in the gaps of the split wires. Such discharge efficiently transforms this uniaxial medium into a conventional wire medium -- that with strong spatial dispersion. Nanoplasma discharges can be generated by an external voltage source.

\section*{Acknowledgements}
This work was supported by the European Innovation Council program Pathfinder Open 2022 through the project PULSE, project number 101099313.

\end{document}